\documentstyle[aps,prl,preprint]{revtex}
\def\beq{\begin{equation}}
\def\eeq{\end{equation}}
\begin{document}                
\title{On the theory of diamagnetism in granular superconductors}
\author{M. V. Feigelman{$^1$} and L. B. Ioffe$^{2,1}$ }
\address{{$^1$}Landau Institute for Theoretical Physics, Moscow, Russia}
\address{{$^2$}Physics Department, Rutgers University, Piscataway, NJ 08855}
\maketitle
\begin{abstract}

We study a highly disordered network of superconducting granules linked by
weak Josephson junctions in magnetic field and develop a mean field theory for
this problem.
The diamagnetic response to a slow {\it variations} of magnetic
field is found to be analogous to the response of a type-II superconductor
with extremely strong pinning.
We calculate an effective penetration depth $\lambda_g$ and critical
current $j_c$ and find that both $\lambda_g^{-1}$ and $j_c$ are non-zero but
are strongly suppressed by frustration.
\end{abstract}
\pacs{}

\narrowtext

In the physics of type II superconductors it is common to distinguish between
weakly and strongly disordered limits of the mixed state.
The former case is described by the flux lattice distorted by
disorder, while in the latter case the vortex loops proliferate and more
adequate description is provided by the model in which the set of
granules is coupled by Josephson junctions.
Whereas the weak disorder limit has been discussed extensively \cite{review},
the opposite limit received much less attention.
It is not a priori clear how these limiting cases are related.
In this paper we apply the methods developed in the theory of spin glasses
and show that response of the Josephson network to the {\em variations} of
magnetic field is qualitatively similar to the response in a weakly
disordered limit.
The goal of this paper is two-fold: it derives the observables characterizing
the Josephson network in strong magnetic field ({\em effective} penetration
depth $\lambda_g$, critical current $j_c$ and critical field $H_c$)
and it establishes parallels between phenomena in a vortex states
(depinning current, shielding, formation of Bean state) and phenomena in
the Josephson networks.

The previous treatment of frustrated Josephson arrays \cite{vinokur},
\cite{ebner} did not consider the spatial (or temporal) variations of magnetic
field making impossible any comparison with the experiment which typically
measures the magnetic moments induced by screening currents.
Here we extend the MF theory so that it includes the magnetic field
variations, the resulting MF equations are analogs of
Ginzburg-Landau equations for the glassy state.
Unlike the Ginzburg-Landau equations which are exact in a wide temperature
range  $G_i \ll (T_c-T)/T \ll 1$ (where Ginzburg parameter $G_i \ll 1$),
the MF equations for the glassy state hold only if
the effective range $\xi_0$ of the Josephson couplings is large compared
to the distance $r_0$ between granules.
This condition is not realized in conventional granular superconductors, but
we hope that results obtained in this approach are still qualitatively
correct and provide reasonable quantitative estimates \cite{hay}.
When deriving the MF equations we shall also assume that the external
magnetic field varies slowly compared to a typical relaxation time of each
junction and that its spatial variations  are small on the scale of $\xi_0$
and that individual loops of Josephson links are too weak to keep the flux
quantum ($\lambda_g \gg \xi_0$).

The analytical results obtained below allow a simple qualitative
interpretation: after cooling in a field below the glass
transition temperature $T_g$ the system is locally trapped
in one of the many metastable states of the formed glassy phase (Fig. 1).
The applied current plays the role of the force acting on a particle in this
relief, since the curvature of each valley is finite, the small applied
force drags the particle only a little up the slope increasing its energy;
it translates to a non-zero $\rho_s$ of a superconducting network.
When the dragging force exceeds a maximal slope of the valley the particle
becomes unstable and crushes to the next valley;
for the superconductor it means that the state with a current exceeding the
critical one becomes unstable, the phase slips and eventually a completely
different state is formed.
Before such instability occurs, the motion is reversible, i.e. the decrease of
the current would lead to the same initial state, implying that the
dissipation is due to these crushes only.
Qualitatively this picture is similar to the vortex pinning and
to the stick and slip model of dry friction.
The role of the generalized force for this state is played by the
variations of the vector potential $\Delta{\bf A}$ (in the gauge
$\nabla{\bf A}=0$), not by magnetic field  because the responses to it
are non-local; in this respect the glassy state behaves as a
superconductor, where the current obeys London equation
${\bf j}=-\rho_s {\bf A}$.

Neglecting the effects of field penetration within each granule and we assume
that each granule is completely specified by the phase of the order parameter
$\phi_i$.
The interaction between granules is due to the Josephson couplings $J_{ij}$:
\beq
H = \frac{1}{2} \sum_{ij} J_{ij} cos(\phi_i- \phi_j - \alpha_{ij})
\eeq
where $\alpha_{ij}= \frac{2\pi}{\Phi_0} \int_i^j {\bf A} d{\bf r}$ is the phase
difference induced by electromagnetic field $\bf A$ and $\Phi_0=\pi\hbar c/e$;
below we shall use the rescaled dimensionless vector potential
${\bf a}=2 \pi \xi_0 \Delta {\bf A}/\Phi_0$ instead of $\Delta {\bf A}$.
It is also convenient to introduce the effective couplings
$\tilde{J}_{ij} = J_{ij}e^{i\alpha_{ij}}$,
then the interaction energy expressed through the variables
$S_i = e^{i \phi_i}$ acquires a form reminiscent of the spin glasses:
$H = \frac{1}{2} \sum_{i,j} J_{ij} S_i^{*} S_j$.
The large background magnetic field $H_0r_0^2 \gg \Phi_0$ ensures that the
couplings $\tilde{J}_{ij}$ are random and correlations between different
couplings are negligible, moreover the variation of the electromagnetic
field $\bf A$ affects the correlation of the effective coupling on the same
link:
\beq
\overline{ \tilde{J}_{ij}(t) \tilde{J}_{kl}(t')}
= \delta_{ik}\delta_{jl} K(r_{ij})[1 -
	(\Delta\alpha_{ij}(t) - \Delta \alpha_{ij}(t'))^2].
\label{JJ}
\eeq
Here $K(r)$ is a smooth function which determines the interaction range
$\xi_0$ via $\xi_0^2 = \frac{1}{d} \int K(r) r^2 d^dr / \int K(r) d^d r$.
In deriving (\ref{JJ})
we assumed that the temporal variation of the induced phase difference
is small: $\Delta \alpha_{ij}(t) \ll 1$ and kept only leading terms in
$\Delta \alpha_{ij}(t) $, because, as we show
below, in the vicinity of $T_c$ the phase slip instabilities start to
occur long before $\Delta \alpha_{ij}(t)$ exceeds unity.

Since the details of the microscopic dynamics are not important at the
macroscopic time scales, we shall assume the simplest overdamped phase
dynamics:
\beq
\frac{1}{\Gamma} \dot{\phi_i} = \sum_j Im \tilde{J_{ij}}e^{i(\phi_i-\phi_j)}
+ \eta \sin\phi_i.
\label{phi}
\eeq
Here the damping coefficient $\Gamma$ can be expressed through the
resistivity $R$ of the individual junction via
$\Gamma \approx \frac{e^2 R}{\hbar^2 n \xi_0^3}$ where $n$ is the number of
granules per unit volume.
In the equations (\ref{phi}) we have introduced an auxiliary local
field $\eta(t)$ which is a convenient tool to probe the phase dynamics.
As we show below, the current induced by the electromagnetic field variation
can be conveniently expressed through the response function
\[
\Delta(t_1,t_2)=\frac{\delta\langle exp[i\phi(t_1)]\rangle}{\delta\eta(t_2)}
	- \tilde{G}(t_1-t_2)
\]
to the field $\eta$ and the correlator
\[
q(t_1,t_2) = \langle exp[i(\phi(t_1)-\phi(t_2))] \rangle - \tilde{C}(t_1-t_2).
\]
Here we define $\Delta(t_1,t_2)$ and $q(t_1,t_2)$ so that they
describe only the long time dynamics of the phase, for this purpose we have
subtracted the short time correlators $\tilde{G}(t_1-t_2)$ and
$\tilde{C}(t_1-t_2)$ decaying at scales $|t_1-t_2| \sim 1/\Gamma$.
With this definition $\Delta(t_1,t_2)$ and $q(t_1,t_2)$ are zero above
the transition temperature where the long term memory is lost,
below $T_g$ $q(t,t)$ coincides with the Edwards-Anderson order parameter.
The appearance of non-zero anomalous response function $\Delta(t_1,t_2)$
below $T_g$ is a hallmark of the glass state, the particular form of both
$\Delta(t_1,t_2)$ and $q(t_1,t_2)$ depends on the preparation of the state,
i.e. on the path in $(T,H)$ plane which lead to the state.
Qualitatively it means that the realized metastable states are different for
different cooling procedures leading to the same final temperature and field.
Note, however, that within the MF approximation the state does not depend
on the cooling rate (as long as it is slow compared to $1/\Gamma$) but is
sensitive only to the path in $(T,H)$ plane.

We shall find that in the model (\ref{JJ}-\ref{phi}) the penetration depth
corresponding to magnetic field {\em variations} is
\beq
\lambda_{g} = \frac{\Phi_0}{4 \pi \sqrt{\pi n T_g}\xi_0} \tau^{-3/2},
\label{lambda_g}
\eeq
where $\tau=(T_g-T)/T \ll 1$;
the linear regime of the field penetration ends for field variations exceeding
\beq
H_c \approx  \upsilon \sqrt{\pi n T_g}\; \tau^{5/2}
\label{H_c}
\eeq
where $\upsilon=0.26$.
At larger field variation the Bean state is formed with the small critical
current
\beq
j_c = \gamma \frac{2 \pi c\xi_0 n T_g}{\Phi_0}\tau^4 \hspace{0.5in}
\gamma=0.065
\label{j_c}
\eeq
The results (\ref{lambda_g}-\ref{j_c}) are the quantitative results of
this letter.
Note that $j_c \sim c H_c/\lambda_g$ as usual.

In order to obtain (\ref{lambda_g}-\ref{j_c}) we shall derive a closed system
of MF equations for the functions $\Delta(t_1,t_2)$ and $q(t_1,t_2)$,
discuss their solutions and finally express $\lambda_g$ and $j_c$
through $\Delta(t_1,t_2)$ and $q(t_1,t_2)$.

The equations for $\Delta(t_1,t_2)$ and $q(t_1,t_2)$ for the infinite range
Josephson network were obtained in Ref.\cite{vinokur} and their analogs for
the Sherrington-Kirkpatrick model in Ref.\cite{ioffe,horner}.
Here we sketch the generalization of the derivation of Ref.\cite{vinokur} for
the finite range model with spatial variations \cite{longpaper}.
We follow the short cut suggested by Kurchan \cite{kurchan} and use the
supersymmetric representation of the spin glass dynamics; in this
approach all correlators are deduced from the disorder-averaged generating
functional $Z=\int \exp(F\{Q\}) {\cal D}Q$:

\begin{eqnarray}
&&F\{Q\} = 2 n T_g \! \int \! d^3{\bf r} \!
\left\{ \left[ \frac{1}{4} \left| [ i\xi_0\nabla +{\bf a}_{2}-{\bf a}_{1}]
	Q(\Theta_1,\Theta_2) \right|^2 +  \nonumber \right. \right. \\
&& \left.
	 \frac{1}{4}(\!\tau_1\!+\!\tau_2\!-\!q_1\!-\!q_2\!)
		Q^2 (\Theta_1,\! \Theta_2) \!+\!
 	 \frac{1}{16} Q^4 (\Theta_1,\! \Theta_2) \right]\!
		  d\Theta_1 d\Theta_2  + \nonumber \\
&& \left.
\frac{1}{6} \int  Q(\Theta_1,\Theta_2)  Q(\Theta_2,\Theta_3)
	Q(\Theta_3,\Theta_1)
	d\Theta_1 d\Theta_2 d\Theta_3 \right\}
\label{F}
\end{eqnarray}
where we denoted $\tau(t_1)=\tau_1$, ${\bf a}(t_1)={\bf a}_1$,
$q(t_1,t_1)=q_1$, etc.
In the functional (\ref{F}) we introduced Grassman coordinates
$\theta$, $\bar{\theta}$ which are supersymmetric partners of the time axis
and used shorthand notations denoting the set $(t,\bar{\theta},\theta)$ by
$\Theta$.
The functional $F\{Q\}$ describes the long term dynamics, so we neglected
in (\ref{F}) the terms containing the time derivatives with coefficients
$\sim 1/\Gamma$.
The superfield $Q(\Theta_1,\Theta_2)$ has many bosonic and fermionic
components, but the only components remaining large in the limit of slow field
variations are bosonic components related to the correlators $\Delta(t_1,t_2)$
and $q(t_1,t_2)$:
\[
Q(\Theta_1,\Theta_2) = q(t_1,t_2) + \bar{\theta}_1 \theta_1 \;
\Delta (t_2,t_1) + \bar{\theta}_2 \theta_2 \;\Delta (t_1,t_2).
\]

The MF approximation is equivalent to a saddle point approximation for the
generating functional $Z$, varying the field $Q(\Theta_1,\Theta_2)$ we
find the MF equations:
\begin{eqnarray}
\begin{array}{c}
[2q^2(t_1,t_2)-\tau^2_1-\tau^2_2-\frac{4}{3}({\bf a}_1-{\bf a}_2)^2]
	\Delta (t_1,t_2) \\
 + \frac{4}{3} \int dt \Delta (t_1,t)\Delta (t,t_2) = 0
\end{array}
\label{Delta_eq} \\
\begin{array}{c}
\lbrack \frac{2}{3} q^2(t_1,t_2)-\tau^2_1-\tau^2_2-\frac{4}{3}
 ({\bf a}_1-{\bf a}_2)^2 \rbrack q(t_1,t_2) \\
 + \frac{4}{3}\int dt (q(t_2,t)\Delta (t,t_1)+q(t,t_1)\Delta (t_2,t)) = 0
\end{array}
\label{q_eq}
\end{eqnarray}
Here the functions $q(t_1,t_2)$, $\Delta(t_1,t_2)$ depend implicitly
on the spatial coordinate $\bf r$ only through the $\bf a(r)$-dependence; this
simplification is
due to the neglect of the gradient terms which is justified in MF
approximation.
The Eq. (\ref{Delta_eq}-\ref{q_eq}) can be transformed to the slow cooling
equations of
infinite range model Ref. \cite{vinokur} by a substitution $a(t) \rightarrow
\delta H(t)$, $\Delta(t_1,t_2) \rightarrow \frac{3}{4}\bar{\Delta}(t_1,t_2)$.

To find the current in a final state we use a general procedure
\cite{longpaper}: we replace
the vector potential ${\bf A}$ by ${\bf A} + {\bf \hat{A}} \bar{\theta}
\theta$ and perform the variational derivative of the generating functional
$j(t) = \left. \frac{\delta Z}{\delta \hat{A}}\right|_{\hat{A}=0}$,
in the MF approximation it simplifies to
$j(t) = \left. \frac{\delta F\{Q\}}{\delta \hat{A}}\right|_{\hat{A}=0}$
evaluated on the saddle point solution $Q(\Theta_1,\Theta_2)$ of equations
(\ref{Delta_eq}-\ref{q_eq}).
Evaluating the derivative we express the current through these solutions:
\beq
{\bf j} = - n\;T_g \frac{4 \pi c\xi_0}{\Phi_0}\int
q(t_1,t_2) \Delta(t_1,t_2)({\bf a}(t_1)-{\bf a}(t_2))\;dt_2
\label{j}
\eeq
The Eqs. (\ref{Delta_eq}-\ref{q_eq}) and (\ref{j}) are analogs of the
Ginzburg-Landau equations for the glassy phase of superconductor.

For a very weak variation of the external field we may replace $\Delta$
and $q$ in this equation by the solution of Eqs. (\ref{Delta_eq}-\ref{q_eq})
in constant field (${\bf a}=0$);
in this case the induced current is related to the vector potential by London
equation:
\beq
{\bf j} = - \rho^{g}_s \delta {\bf A}
\label{rho_g}
\eeq
\[
\rho^{g}_s = \frac{8 \pi^2 c \xi_0^2 n T_g}{\Phi_0^2}
	\int q_0(t_1,t)\Delta_0(t_1,t)dt =
\frac{4 \pi^2c\xi_0^2 n T_g}{\Phi_0^2}
	\tau^3
\]
Here we used the exact solution ($\Delta$, $q$) of the
(\ref{Delta_eq}-\ref{q_eq}) for
the cooling at constant field \cite{vinokur}.
Thus, the superconductor in the glassy state shields the {\em variation}
of the magnetic field; using the superfluid density (\ref{rho_g})
we find $\lambda_g$ (\ref{lambda_g}).

The solution of the equations (\ref{Delta_eq}-\ref{q_eq}) changes completely
when $a$
becomes of the order of $\tau$, in this regime we must solve these non-linear
equations from scratch and evaluate current using (\ref{j}).
Dimensional analysis shows that for the field variation that takes place at
constant temperature the current is
\beq
{\bf j}({\bf a}) = - {\bf e_a} \frac{2 \pi c \xi_0 n T_g}{\Phi_0} \tau^4 \cdot
	\Upsilon \left( \frac{a}{\tau} \right)
\label{j2}
\eeq
Here ${\bf e_a} = {\bf a}/|a|$ and the dimensionless function
$\Upsilon(\alpha)$
was evaluated numerically, we show its plot in Fig 2; note that
$\Upsilon(\alpha) =\alpha$ at small $\alpha \ll 1$, reproducing (\ref{rho_g}).

The current reaches the maximum value for $a_c=0.18 \tau$, we were not able to
find the stable numerical solution of the equations
(\ref{Delta_eq}-\ref{q_eq}) for $a \geq a_c$.
As long as the solution of the equations (\ref{Delta_eq}-\ref{q_eq}) exists,
the changes induced by the variations of $a$ are reversible, i.e.
if one changes the field $a$ from its initial zero value at $t_0$ to
$a(t_1)$, then to the maximal value $a(t_m)$ and then back to
$a(t_f)= a(t_1)$, the current (and all other properties) in the
final state are the same as at $t_1$.
To prove it we define function $\vartheta(t)$ so that $a(\vartheta(t))=a(t)$
and $\vartheta(t) \leq t_m$ (note that with this definition $\vartheta(t)=t$
at $t\leq t_m$).
Suppose that $\Delta_0(t_1,t_2)$, $q_0(t_1,t_2)$ are solutions of
(\ref{Delta_eq}-\ref{q_eq}) for the monotonic variation of the field ($t_1,t_2
\leq t_m$), then the solution in the whole
range $t_1,t_2 \leq t_f$ is given by
\beq
\begin{array}{r}
\Delta(t_1,t_2) = \theta(t_1-t_2) \Delta_0(\vartheta(t_1),\vartheta(t_2) \\
q(t_1,t_2)=q_0(\vartheta(t_1),\vartheta(t_2))
\end{array}
\label{ansatz}
\eeq
which is verified by the substitution into (\ref{Delta_eq}-\ref{q_eq}).
Inserting the solution (\ref{ansatz}) into (\ref{j}) we see that
$j(\vartheta(t))=j(t)$ proving the reversibility of small variations of
$a \leq a_c$.

Clearly, large variation of magnetic field should eventually lead to the
current dissipation and a loss of reversibility.
Thus, the solution of the Eqs.~(\ref{Delta_eq}-\ref{q_eq}) can not exist
beyond some critical value of $a$.
This can be  proved formally, the proof goes in three steps: (i) for a
large constant variation of the magnetic field all memory of the initial
couplings $J_{ij}(t_0)$ is lost (see (\ref{JJ})), so the order parameters
depend only on the time difference: $\Delta(t_1-t_2)$, $q(t_1-t_2)$.
Now we find the asymptotics of these functions as $t \rightarrow \infty$;
(ii) we prove by the direct substitution into Eq.~(\ref{Delta_eq}) that
solution which is identically zero at large times
($\Delta(t) > 0$ at $t<t_b$, but  $\Delta(t)= 0$ at $t>t_b$) is impossible;
and finally (iii) we show that
at $t \rightarrow \infty$ Eq.~(\ref{Delta_eq}) does not admit
decreasing solution (instead it gives $\Delta(t) \propto t$ which make no
sense).
Unfortunately, this does not establish the maximal value of $a$ which admits
the solution of Eqs.~(\ref{Delta_eq}-\ref{q_eq}), but it is likely
that this value coincides with $a_c$ at which
the current is maximal and numerical
stability of the solution of (\ref{Delta_eq}-\ref{q_eq}) is lost.

The Eqs.~(\ref{Delta_eq}-\ref{q_eq}) were derived under the assumption that
the variations of the phase variables happens on the same large time scales as
the variation of the external field.
The absence of the solution of these equations can only mean that this
assumption no longer holds and a fast phase slip occurs at $t_c$ when
$a(t_c)=a_c$.
The state forming at $t>t_c$ has no correlations with the state
before the catastrophe, this state resembles the nascent state obtained
after a new fast cooling.
Since the rate of cooling does not affect the forming state as long as this
rate is slow compared with microscopic times scales, the nascent state can be
described by the solutions of the slow cooling
equations  ~(\ref{Delta_eq}-\ref{q_eq}) in which the initial period of slow
cooling in a constant magnetic field $t_0>t>0$ was compressed to a infinitely
narrow time interval $(t_c,t_c+\epsilon)$.
The solution of Eqs.~(\ref{Delta_eq}-\ref{q_eq}) for such cooling is related
to the solution for unit rate cooling by rescaling:
$\Delta(t_1,t_2)=\Delta_0(t_1-\tilde{t_c},\tau(t_2)) \frac{d\tau}{dt_2}
\approx \tilde{\Delta}(t_1)\delta(t_2-t_c)$ for $t_c < t_2 < t_c+\epsilon$
where $\tilde{t_c}=t_c-t_0$,
$\tilde{\Delta}(t_1)=\int_{0}^{t_0}\Delta_0(t_1-\tilde{t_c},t_2) dt_2$ and
$\Delta_0(t_1,t_2)$, $q_0(t_1,t_2)$ is the solution for the initial period
($t_c>t_1>0$ and $t_c>t_2>0$) of the field variation.
Therefore the solution after the phase slip is given by
\[
\begin{array}{ll}
\Delta(t_1,t_2)=& \theta(t_1-t_c) \theta(t_2-t_c) \Delta_0(t_1-\tilde{t_c},
	t_2-\tilde{t_c})+\\
	& \tilde{\Delta}(t_1)\delta(t_2-t_c) \\
q(t_1,t_2)=& \theta(t_1-t_c) \theta(t_2-t_c) q_0(t_1-\tilde{t_c},
	t_2-\tilde{t_c})
\end{array}
\]
The continuing the field variation at $t>t_c$ results in a current growing
as ${\bf j}({\bf a}\! -\! {\bf a}_c)$ until the variation of $a$
exceeds $2a_c$, then the second phase slip occurs and so on.

For a very small variation of the external field the screening current falls
off exponentially inside the sample: $j = j(0) \exp(-\lambda_g z)$.
For larger fields $j(0)$ becomes comparable with $j_c$ and the non-linear
differential equation
$\frac{d^2 {\bf A}}{dz^2} = \frac{4\pi}{c} {\bf j}\left(
\frac{2\pi \xi_0 \bf A}{\Phi_0} \right)$
should be solved to find the current profile.
At even higher fields $\Delta H > H_c$ the current at the boundary exceeds
$j_c$ and
the dissipation begins, the field $H_c$ is somewhat analogous to a critical
field of a type II superconductor, but here the fields ($H_{c1}$) for
which vortices penetrate into the sample and the fields ($H_c$) that produce
the pair breaking current coincide.
The vector potential corresponding to $H_c$
equals $\Phi_0a_c/(2\pi\xi_0)$ at the boundary and decays on scales
$\lambda_g$, so
$H_c \approx \frac{\Phi_0a_c}{2\pi\xi_0\lambda_g}$; to find the numerical
coefficient we solved numerically the equation for the current profile and got
(\ref{H_c}).

In this letter we did not consider two important physical effects present in
real networks: (i) the magnetic field is shielded by individual
granules leading to field inhomogeneity at small scales and
(ii) exponentially rare transitions over the energy barriers missed by MF.
It is not clear to us how large are these effects in real
materials.

Experimentally, the study of Josephson networks is complicated by the
difficulty of separating the response of the individual granules
from the collective effects of Josephson currents.
Recently Ocio and Leylekian have surpassed this difficulty \cite{ocio}
preparing the state by cooling the sample in a large field and then
studying the responses to a very small variations of field and temperature.
They found that the reaction of this state to the {\em change} of
magnetic field $\Delta H$ can be characterized by large $\lambda_g$ and a
small, but non-vanishing, $j_c$, e.g. they found a complete shielding of
magnetic field {\em variations}.

In conclusion we found that the effect of the variation $\Delta H$ of the
external magnetic field on the Josephson network cooled in a constant
magnetic field is qualitatively similar to the effect of the external field
on the type II superconductor with an extremely strong pinning.
Quantitatively, however, the superconductivity of the Josephson network is
much weaker: the expressions for penetration depth, critical current,
critical field contain large powers of reduced temperature and small numerical
factors.
For instance, $\frac{\lambda_g}{\lambda_0}=\sqrt{\frac{T_c}{T_g}}
\frac{2}{\tau}$, $\frac{j_c^{g}}{j_c^0}=\frac{3\sqrt{3}\gamma}{8}
\frac{T_g}{T_c} \tau^{5/2}$ where $\lambda_0$, $j_c^0$ and $T_c$ are the
penetration depth, critical current and transition temperature of the same
network in zero external magnetic field and we compare properties for the same
reduced temperatures $\tau$.

We are grateful to  L. Leylekian and M. Ocio  for numerous inspiring
discussions and  to P. Chandra for useful critique.
This research was supported by the innovation partnership grant of NJ and by
ISF through grant M6M000.

\begin{figure}
\caption{A typical energy relief of the glassy state; before (full line) and
after (dashed line)
the magnetic field was changed. Arrows show irreversible processes.}
\end{figure}

\begin{figure}
\caption{Induced scaled current $\Upsilon(\alpha)$ as a function of the vector
potential variations $\alpha=a/\tau$}
\end{figure}

\end{document}